\documentclass[twocolumn,aps]{revtex4}

\usepackage{graphicx}

\begin{document}

\preprint{XXX}

\title{Impact of the Casimir-Polder Potential and Johnson Noise on Bose-Einstein Condensate Stability near Surfaces}

\author{Yu-ju Lin, Igor Teper, Cheng Chin, and Vladan Vuleti\'{c}}

\affiliation{Department of Physics, Stanford University, Stanford,
California 94305-4060}

\altaffiliation{Present Address: MIT-Harvard Center for Ultracold
Atoms, MIT, Cambridge, MA 02139.}

\date{\today}

\begin{abstract}
We investigate the stability of magnetically trapped atomic
Bose-Einstein condensates and thermal clouds near the transition
temperature at small distances 0.5 $\mu$m $\leq d \leq$ 10 $\mu$m
from a microfabricated silicon chip. For a 2 $\mu$m thick copper
film the trap lifetime is limited by Johnson-noise induced
currents and falls below 1 s at a distance of 4 $\mu$m. A
dielectric surface does not adversely affect the sample until the
attractive Casimir-Polder potential significantly reduces the trap
depth.
\end{abstract}

\maketitle

Bose-Einstein condensates in magnetic traps
\cite{Haensel01,Ott01,Leanhardt02,Leanhardt03,Schneider03} and
waveguides \cite{Leanhardt02,Fortagh03}, and thermal atoms in
waveguides \cite{Mueller99} produced by microfabricated structures
(microtraps) hold great promise for new quantum devices for atomic
matter waves, such as Fabry-P\'{e}rot resonators \cite{Wilkens93},
interferometers \cite{Shin03}, or Josephson junctions
\cite{Anderson98}. Full quantum control over the motion of an
ultracold atom of mass $m$ and energy $E$ requires potentials that
vary abruptly on a length scale $\lambda \sim$ $h/(mE)^{1/2} \sim$
1 $\mu$m. Such potentials can be created at small distances
$d\leq\lambda$ from miniaturized field sources.

However, the proximity of a room-temperature surface can perturb
the ultracold gas, and microtrap experiments have revealed
condensate fragmentation \cite{Fortagh02,Leanhardt02,Leanhardt03},
heating \cite{Haensel01,Fortagh02}, and reduced trap lifetime
\cite{Fortagh02}. The fragmentation has been traced to spatial
variations of the longitudinal magnetic field near a conductor
carrying current \cite{Kraft02}, while heating and loss have been
eliminated for distances $d \geq 70$ $\mu$m by careful electronic
design and shielding \cite{Leanhardt03}. However, Jones {\it et
al.} \cite{Jones03} and Harber {\it et al.} \cite{Harber03} have
recently reported a fundamental limit due to spin flips induced by
thermally excited currents in a mesoscopic conductor, in very good
agreement with theoretical predictions \cite{Henkel99a} over the
measurement regions 25 $\mu$m $\leq d \leq$ 100 $\mu$m and 3
$\mu$m $\leq d \leq$ 1 mm, respectively.

In this Letter, we explore fundamental limitations on condensate
stability at small distances down to $d= 0.5$ $\mu$m from
dielectric and metal surfaces. For a 2 $\mu$m thick copper film
carrying no current we observe a distance-dependent lifetime
$\tau(d)$ that is quantitatively explained by thermal magnetic
field fluctuations arising from Johnson-noise induced currents
\cite{Varpula84,Henkel99a}. For the dielectric, we observe a
reduction in trap lifetime only when the vicinity of the surface
limits the trap depth. A one-dimensional (1D) evaporation model
can explain the measured trap loss, but only when the attractive
Casimir-Polder force \cite{Casimir48} between atoms and surface is
included. Our results suggest that the local manipulation of
condensates will be possible using thin conductors, which have low
magnetic field noise, on dielectric surfaces.

\begin{figure}
\includegraphics[width=3.2in]{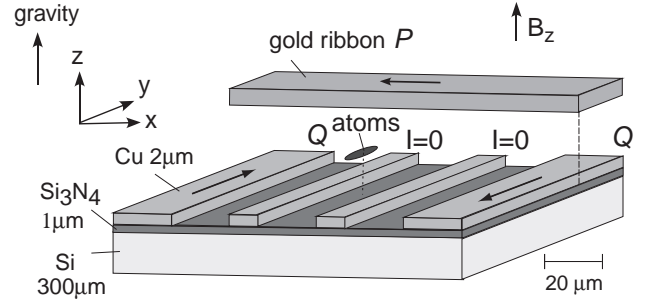}
\caption{Microfabricated chip. (Only the $x$ direction is to
scale.) The outer Cu wires ($Q$) generate a 2D quadrupole field in
the $xz$ plane. The ribbon ($P$) in combination with an external
field gradient creates the confinement along $y$.}\label{setup}
\end{figure}

The atom-surface interactions are studied using ultracold atoms
confined in a Ioffe-Pritchard trap generated by currents flowing
in microfabricated conductors on a silicon chip (see
Fig.~\ref{setup}). The chip was produced by first coating a 300
$\mu$m thick silicon substrate with a 1 $\mu$m thick, electrically
insulating Si$_3$N$_4$ diffusion barrier using plasma-enhanced
chemical vapor deposition. Subsequently, a 20 nm thick Ti adhesion
layer, a 2.15(20) $\mu$m thick Cu conducting layer, a 40 nm thick
Ti, a 50 nm thick Pd, and a 100 nm thick Au layer were deposited
by electron beam evaporation. Finally, the wires were defined by
photolithography and wet etching.

The radial ($xz$) confinement of the Ioffe trap formed above the
chip is provided by a 2D quadrupole field, generated by two copper
wires $Q$ along the $y$ direction carrying antiparallel currents,
in superposition with a bias field along $z$. The centers of the 2
$\mu$m thick and 20 $\mu$m wide $Q$ wires are separated by 100
$\mu$m. The axial ($y$) confinement is created by a
current-carrying gold ribbon $P$ along $x$ in combination with an
external field gradient along $y$. The 25 $\mu$m thick, 150 $\mu$m
wide ribbon 155 $\mu$m above the chip was wire-bonded to the
surface.

\begin{figure}
\includegraphics[width=3.3in]{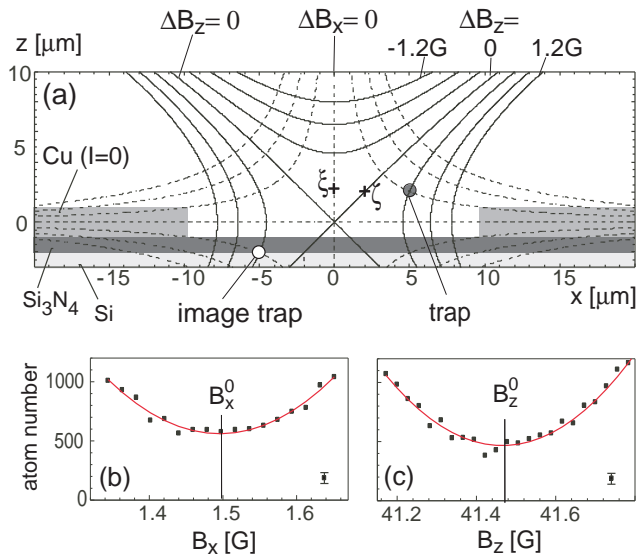}
\caption{(a) Map of relative field ($\Delta B_x,\Delta B_z$) to
trap position in the $xz$ plane. Dashed (solid) lines are contours
of $\Delta B_x$ ($\Delta B_z$), plotted with 0.4 G spacing. (b,c)
Measured atom loss versus $B_x$ ($B_z$) near point $\xi$
($\zeta$), marked by a cross. } \label{contour}
\end{figure}

The condensate production starts with a standard magneto-optical
trap (MOT), into which typically $10^7$ $^{87}$Rb atoms are
collected within 8 s from a Rb dispenser beam \cite{Ott01}. We
then move the MOT cloud from the original distance $d=17$ mm to
within $d=6$ mm of the chip surface, and compress it for 20 ms.
After the MOT light has been extinguished, the atoms are optically
pumped in 400 $\mu$s into the $F = 2$, $m = 2$ ground state, and
loaded into a large quadrupole trap with $x,y,z$ gradients of 33,
8, and 25 G/cm, respectively. Next, by increasing a bias field
$B_z$ generated by a small coil located 2 mm below the chip, the
atoms are compressed in 300 ms into a Ioffe microtrap, where the
conversion from quadrupole to Ioffe trap geometry follows Ref.
\cite{Fortagh98}. The trap is located at $d=50$ $\mu $m from the
surface and 430 $\mu $m away from the $P$ ribbon along $y$ for
currents of 0.6 A in both the $Q$ wires and the $P$ ribbon. For a
field at trap bottom of $B_0$=1.5 G, the radial and axial
vibration frequencies are $\omega_{rad}/2 \pi$=5.1 kHz and
$\omega_{ax}/2 \pi$=70 Hz, respectively. We typically load $3
\times 10^6$ atoms at an initial temperature of 300 $\mu$K, peak
density of $1.7 \times 10^{12}$ cm$^{-3}$, and peak collision rate
of 140 s$^{-1}$. Three seconds of forced evaporation cool the
sample to below the transition temperature $T_c=0.8$ $\mu$K. When
the thermal component is no longer discernible in a time-of-flight
image, the condensate contains $10^3$ atoms at a peak density of
$n_p = 8 \times 10^{14}$ cm$^{-3}$. To measure surface-induced
loss, we transport the condensate or a cloud near $T_c$
adiabatically in 40 ms to a defined position near the surface,
hold it there for a variable time, and image the cloud after
moving it back to $d=100$ $\mu$m. The
noise-equivalent-optical-density of 1\% in the absorption imaging
corresponds to a small atom number noise $\delta N$= 50 for a
condensate. The procedure is repeated for each parameter value.

In order to compare an observed influence of the surface to
theoretical models, the accurate calibration of the trap position
($x$,$z$) in the $xz$ plane is crucial. While optical imaging
fails close to the chip, the calibration is facilitated by the
symmetry of the photolithographically defined $Q$ wires (see
Fig.~\ref{setup}). The trap is located where a homogeneous
external bias field, with $x$ and $z$ components ($B_x$,$B_z$),
cancels the field from the $Q$ wires. Once the bias field value
($B_x^0$,$B_z^0$) that places the trap at the symmetry center
($x$=0,$z$=0) of the $Q$ wires is known precisely, the atoms can
be accurately positioned at arbitrary ($x$,$z$) by applying an
additional field $(\Delta B_x,\Delta B_z)=(B_x-B_x^0,B_z-B_z^0)$
to compensate the spatially varying field from the $Q$ wires.
Fig.~\ref{contour}a shows how the relative field ($\Delta
B_x$,$\Delta B_z$) maps to trap position ($x$,$z$). We also take
into account a slight map distortion due to all other coils in the
setup, which displaces the Ioffe trap by $0.50(5)$ $\mu$m away
from the chip compared to the map defined by the $Q$ wires alone.

To precisely measure the symmetry-center bias field
($B_x^0$,$B_z^0$) that depends on unknown stray fields, we make
use of the reflection symmetry of the $Q$ wire configuration. We
exploit the fact that a mirror image trap, located at ($-x$,$-z$),
coexists with the trap at ($x$,$z$) (see Fig.~\ref{contour}a). As
the trap and the image trap are brought close, the atoms can
overcome the barrier between the traps, and will be lost if the
image trap is inside the surface. Along a $\Delta B_z$ contour,
the loss is symmetric about the point $\Delta B_x=0$ (e.g., point
$\xi$ in Fig.~\ref{contour}a), where the minimum barrier leads to
maximum loss. From the measured atom number versus $\Delta B_x$
along the $\Delta B_z$= -120 mG contour, we determine $B_x^{0}$
with a precision of $\delta B_x^0$= 4 mG (Fig.~\ref{contour}b).
Similarly, $B_z^{0}$ is determined within $\delta B_z^0$= 10 mG
(Fig.~\ref{contour}c) by a measurement along the $\Delta B_x$= 110
mG contour near point $\zeta$ in Fig.~\ref{contour}a. In the
spatial region of interest, the uncertainties $(\delta
B_x^0,\delta B_z^0)$ correspond to a trap $z$ position error of 20
nm, small compared to a condensate size of 300 nm in the $xz$
plane.

\begin{figure}
\includegraphics[width=3.3in]{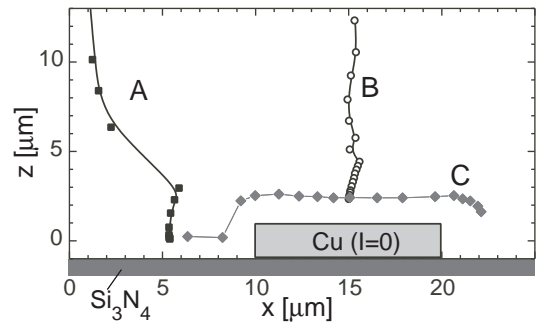}
\caption{Paths chosen for trap lifetime measurements above a
dielectric surface ($A$) and above a copper film ($B$). Line $C$
is the measured contour line of 22 ms lifetime near the
metal.}\label{path}
\end{figure}

To verify that we can position the trap accurately relative to
microscopic structures on the surface, we first measure a line of
constant lifetime $\tau=22$ ms near a Cu film carrying no current,
shown as curve $C$ in Fig.~\ref{path}. This ``surface microscopy''
yields a contour that displays the expected symmetry about the
metal, which confirms the skewed field-to-position mapping in
Fig.~\ref{contour}a. Fig.~\ref{path} also shows the trajectories
we then use to measure the lifetime above the Si$_3$N$_4$
dielectric ($A$) and the copper film ($B$). Path $A$ is selected
to avoid coupling to the image trap in the center region.

\begin{figure}
\includegraphics[width=3.3in]{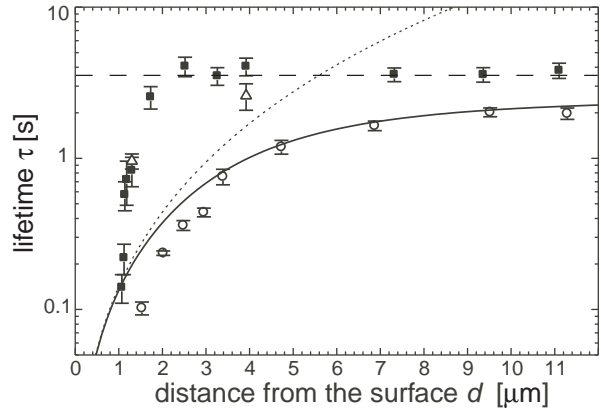}
\caption{Trap lifetime as a function of distance from a dielectric
(solid squares) and metal (open circles) surface, for $T$=1 $\mu$K
and $B_0$=0.57 G. The dotted line is the calculated lifetime above
the metal due to thermal $B$ fields only, the solid line includes
the one-body lifetime. The open triangles are measurements for a
pure condensate above the dielectric.} \label{lifetime}
\end{figure}

Fig.~\ref{lifetime} shows the lifetime $\tau$ as a function of
distance $d$ from the respective surface, measured at $T=1$ $\mu$K
($T/T_c=1.3$) for an offset field $B_0$= 0.57 G. The lifetime
above the dielectric is constant for $d \geq 2.5$ $\mu$m, while
above the metal $\tau$ is shorter and distance-dependent. Since
even a conductor carrying no macroscopic current generates
magnetic field fluctuations associated with thermal current noise
\cite{Varpula84}, it can induce trap decay by driving transitions
from trapped to untrapped atomic sublevels \cite{Henkel99a}. In
the limit that the metal film thickness (here $t$=2.15(20) $\mu$m)
is much smaller than the skin depth $\delta$ at the transition
frequency ($\delta$=103 $\mu$m for $B_0 = 0.57$ G), the current
noise is the Johnson noise in the conductor, and is
frequency-independent. (The measurements in Refs.
\cite{Jones03,Harber03} were performed in the opposite limit, $t
\gg \delta$, of a bulk metal.) Then for a metal film of width $w
\gg t$, and resistivity $\rho$ at temperature $T$, the spin flip
rate $|F,m \rangle \rightarrow |F,m-1 \rangle$ is given by
$\Gamma_{Fm}= C_{Fm}^2 C_0[d(1+d/t)(1+2d/w))]^{-1}$. This formula
interpolates the loss rates predicted by Henkel {\it et al.}
\cite{Henkel99a} in the limits $d \ll w$ and $d \gg w$. We derive
$\Gamma_{Fm}$ at $d \gg w$ from Ref. \cite{Henkel99a} assuming
only thermal currents along the wire contribute substantially.
Here $C_{Fm}^2= | \langle F,m-1| S_- |F,m \rangle |^2$, $S_-$ is
the electron spin lowering operator, $C_0= 88$ s$^{-1} \mu$m
$\times (T$/300K)$(\rho_{Cu}/\rho)$, $\rho_{Cu}=1.7 \times
10^{-8}$ $\Omega$m, $T$ = 400 K from the measured
$\rho/\rho_{Cu}$, and $w$=10 $\mu$m. We assume the atoms are lost
in a cascade process, $|2,2 \rangle \rightarrow|2,1 \rangle
\rightarrow|2,0 \rangle$, replace $C_{Fm}^2$ by ($C_{22}^{-2}+
C_{21}^{-2})^{-1}=(4+\frac{8}{3})^{-1}$, and add the
distance-independent one-body loss rate $\gamma_1 = 0.4$ s$^{-1}$
observed at $d \geq$ 10 $\mu $m. The result (solid line) agrees
well with the observed lifetime above the thin copper film. For
comparison, the fundamental limit due to thermal field noise only
($\gamma_1=0$) is plotted as a dotted line. Except for the point
closest to the metal surface, $\tau$ is independent of sample
temperature, indicating that the loss process is not evaporation
at finite trap depth. Further, the lifetime $\tau(d)$ measured for
$B_0$= 1.5 G, i.e. at a three times larger transition frequency,
is compatible with white field noise within 40\%.

\begin{figure}
\includegraphics[width=3.3in]{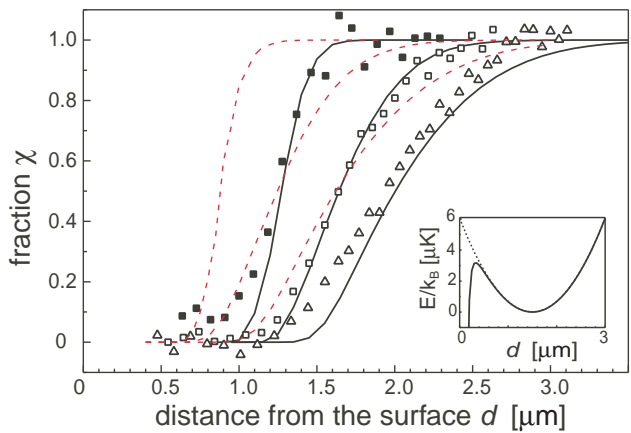}
\caption{Remaining atom fraction $\chi$ in a trap at distance $d$
from dielectric surface for a condensate (solid squares), and for
thermal clouds at 2.1 $\mu$K (open squares) and 4.6 $\mu$K
(triangles). The solid (dashed) lines are calculated with
(without) Casimir-Polder potential for the BEC, 2.1 $\mu$K, 4.6
$\mu$K clouds (left to right). The inset shows the trapping
potentials for $C_4 = 8.2 \times 10^{-56}$ Jm$^4$ (solid line) and
$C_4 = 0$ (dotted line).} \label{casimir}
\end{figure}

Above the dielectric, the constant lifetime $\tau$=3.5 s observed
for $d>2.5$ $\mu$m is independent of cloud temperature for 1
$\mu$K$\leq T \leq$3 $\mu$K, and the latter remains constant
within our resolution of 0.25 $\mu$K/s. In the short-distance
region of decreasing lifetime, however, $\tau$ is longer for a
colder cloud, which is consistent with surface-induced 1D
evaporation \cite{Reichel99,Harber03}. To test this explanation,
we measure the remaining atom fraction $\chi$ after 15 ms versus
$d$ for a condensate, and for thermal clouds at 2.1 $\mu$K and 4.6
$\mu$K (Fig.~\ref{casimir}). A thermal cloud exhibits loss at a
larger distance than a condensate, and the latter vanishes at a
finite distance $d=1$ $\mu$m.

In the absence of atom-surface interactions, the trap depth would
be given by the value of the trapping potential at the surface.
However, as shown in the inset to Fig.~\ref{casimir}, the
attractive Casimir-Polder potential \cite{Casimir48},
$V_{CP}=-C_4/d^4$, lowers the trap depth, and the trap disappears
at finite $d$. To quantify this effect, we model the process as a
sudden loss of the Boltzmann tail as the atoms are brought near
the surface, in conjunction with 1D evaporation for $t_0=15$ ms in
a trap with $\omega_{rad}/ 2 \pi=3.6$ kHz. The remaining fraction
after the sudden loss is given by $F=1-e^{-\eta}$, where $\eta =
U_0 /(kT)$ is the ratio of the Casimir-force limited trap depth
$U_0(d)$ and thermal energy $kT$. We account for atom tunneling
through the barrier as a small reduction in the effective trap
depth. The loss rate for 1D evaporation is $\Gamma = f(\eta)
e^{-\eta} /\tau_{el}$, which using $f(\eta) =
2^{-5/2}(1-\eta^{-1}+\frac{3}{2}\eta^{-2})$ is accurate to 5\% for
$\eta \geq$ 4 \cite{Ketterle96}. Given elastic collision times
$\tau_{el}=$ 0.2 ms, 0.9 ms, and 1.5 ms for the condensate, the
2.1 $\mu$K and the 4.6 $\mu$K clouds, respectively, we plot in
Fig.~\ref{casimir} the remaining fraction $\chi_{CP}=F e^{-\Gamma
t_0}$. For the condensate, we assume that the loss is due to
collisions between a residual thermal cloud at $T_c/2$ and the
condensate. For the Casimir potential we use the coefficient $C_4
= \psi(\varepsilon) 3\hbar c \alpha / (32 \pi^{2} \epsilon_{0})$,
where $\alpha = 5.25 \times 10^{-39}$ Fm$^{2}$ is the Rb
polarizability, and $\psi(\varepsilon)=0.46(5)$ is a numerical
factor from Ref. \cite{Yan97} for the Si$_3$N$_4$ dielectric
constant of $\varepsilon$ = 4.0(8). For comparison, we also plot
the calculated fraction in the absence of any surface potential
($C_4=0$).

Fig.~\ref{casimir} can be interpreted as a measurement of the
Casimir-Polder coefficient $C_4$ that is, however, limited by the
uncertainty of the distance calibration. The dominant contribution
of $\pm$100 nm arises from a $\pm$200 nm uncertainty of the
conductor thickness $t$ (see Fig.~\ref{contour}). In addition, an
estimated 10\% field calibration uncertainty for $\Delta B_x$
contributes a 10\% scaling error about the distance $d_0=1.6$
$\mu$m. Furthermore, our setup with a 1 $\mu$m Si$_3$N$_4$ layer
on Si is a dielectric waveguide. The corresponding correction to
$C_4$ compared to a Si$_3$N$_4$ half-space is estimated to be less
than 20\% \cite{Courtois96}. Uncertainties in temperature and trap
vibration frequency of 10\% also affect $\chi_{CP} (d)$. When we
take these error sources into account, our measurements in
combination with the 1D evaporation model imply a 66 \% confidence
range for $C_4$ between $1.2 \times 10^{-56}$ Jm$^4$ and $41
\times 10^{-56}$ Jm$^4$, which includes the nominal value $C_4 =
8.2 \times 10^{-56}$ Jm$^4$. The good agreement between our data
and the predicted $C_4$ without any adjustment of parameters
suggests that the Casimir potential limits the trap depth, and
consequently lifetime, at small distances $d \leq 2$ $\mu$m from
the dielectric surface. The discrepancy at small fraction $\chi$
is probably due to our simple 1D evaporation model which breaks
down for $\eta \leq 1$, and also ignores evaporation-induced
temperature changes. Our data exclude $C_4=0$, even if we allow
for the largest possible systematic error.

In conclusion, we have characterized the stability of magnetically
trapped ultracold atoms at $\mu$m distances from a copper film and
a dielectric surface. The condensate is stable over the
dielectric, and the spectral density of the thermal magnetic field
near a metal film scales with metal thickness. Therefore it will
be possible to bring a stable ultracold cloud sufficiently close
to the surface for the trapping potential to be locally
manipulated.

This work was supported by the ARO. We thank M. Kasevich for
stimulating discussions and X. Wu for technical assistance.

{\it Note added }: The Casimir potential cannot explain the
anomalously short lifetime at the smallest distance $d=1.5$ $\mu$m
above the metal (see Fig.~\ref{lifetime} and the ``surface
microscopy'' curve $C$ in Fig.~\ref{path}). One possible
explanation is patch potentials from Rb atoms adsorbed on the
metal, as recently reported by McGuirk {\it et al.}
\cite{McGuirk03}.


\end{document}